**Electronic density of states as the descriptor of elastic bond strength, ductility, and local lattice distortion in BCC refractory alloys**

[1]Dharmendra Pant and [1]Dilpuneet S. Aidhy

[1]Department of Materials Science and Engineering, Clemson University, SC 29634, USA

Corresponding author: daidhy@clemson.edu

## Abstract

Although electronic density of states (DOS) is fundamental to materials properties, its general relationship to mechanical properties of alloys is not well established. In this paper, using density functional theory (DFT) calculations, we show that the electronic occupancy at the Fermi level, N($E_f$), obtained from DOS is a key descriptor of alloy strength and ductility. Our comprehensive analysis of numerous body centered cubic (BCC) refractory high entropy alloys (RHEAs) shows an overwhelming correlation that low N($E_f$) indicates strong bonds that have high stiffness resulting in high elastic constants. High bond stiffness indicates presence of covalent nature of bonds that are directional in nature resulting in resistance to deformation leading to high bulk ($B$) and shear ($G$) moduli. Consequently, N($E_f$) provides a direct correlation to the tendency of alloy ductility evidenced in the Pugh ratio ($G$/$B$). As stiffer bonds result in lower local lattice distortion (LLD), N($E_f$) are LLD are also found to be corelated which opens up a correlation to solid solution strengthening and yield strength. Thus, this work unveils fundamental correlations between N($E_f$) and (1) elastic bond strength, (2) ductility, and (3) LLD. These correlations open opportunities for the design of high strength high ductile RHEAs.

## 1. Introduction

The atomic and electronic level properties of materials are anchored in the atomic bonding that is described by the electronic charge distribution. The electronic density of states (DOS) is widely used to determine electronic properties tied to energy states such as band gap and reaction energetics, and a profound understanding of the correlations between electronic properties and DOS exists. The electronic structure is also fundamental to bond strength and cohesive energy of materials that are responsible for mechanical properties. However, there is a limited understanding that bridges DOS and mechanical properties, especially in complex, chemically-random concentrated alloys. This knowledge is under intense pursuit for the development of high-strength, high-ductility refractory high entropy alloys (RHEAs) for high temperature structural applications[1–8]. RHEAs are a new class of high-temperature structural materials, mainly composed of Cr, Mo, W, V, Nb, Ta, Ti, Zr and Hf, have gained significant attention due to their excellent mechanical properties, including high strength, thermal stability, and oxidation resistance[9,10], which contribute to their superior performance compared to conventional alloys. However, RHEAs face critical challenges, such as intrinsic brittleness at room temperature[11], limited plasticity[12], and difficulties in balancing strength and ductility[13]. These issues can be addressed through strategies such as phase engineering[14], microstructure optimization[15], valence electron optimization [16,17], metastability-engineering[18] and computational alloy design. Although current efforts to overcome these challenges in RHEAs are promising, a deeper understanding of the electronic structures of these alloys and their correlation with mechanical properties has now become crucial.

The valence electronic configuration (VEC) has been used as the electronic descriptor to capture the role of electronic structure, especially in estimating the crystal structure and phase stability[19–22]. Recent works have also probed correlations between VEC and ductility parameters but with limited success[2,23,24]; the correlation is weak especially as the number of elements increase in a given composition. Almost three decades prior, investigations were performed where the relative stability among crystal structures in intermetallics of $Ni_3X$ precipitates were corelated to the electronic occupancy at the Fermi level, $N(E_f)$, in Ni-based superalloys[25–31]. In addition, some trends that related the bond strength, $N(E_f)$, and elastic constants were also identified[29]. In this work, we extend this knowledge to HEAs and show that $N(E_f)$ is a key descriptor for both strength and ductility in RHEAs.

In a classic text-book description, the metallic, ionic and covalent bonds are described by the 'sea-of-electrons', electron-transfer, and electron-sharing, respectively. The resulting charge distribution determines the electronic states' occupancy and gives rise to band gap in most covalent and ionic materials with $N(E_f)=0$. These materials have stiffer bonds attributed to higher charge transfer; they are therefore generally stronger. In contrast, the metals have non-zero electronic occupancy at the Fermi level. The bonds in metals are less stiff and lack directionality, resulting in lower strength. The DOS of the *d*-electrons in refractory metals is described by the pseudo-gap at the Fermi level, often evaluated by *d*-band bimodality, and previous works have probed the *d*-band filling trends in the different transition metals[1,29,32]. In this paper, we show that N($E_f$) is directly correlated to the bond strength in metals, i.e., lower the occupancy, higher is the strength. The strength of metals and alloys directly scales with N($E_f$). This relationship consolidates anecdotal recent evidence in HEA literature. For example, Kaneko and Yoshimi[32] posit that $Mo_xCr_{1-x}$ is stronger than $Mo_xTi_{1-x}$ due to lower N($E_f$) in the former. Similarly, Startt et al.[33] showed that increasing Mo content in MoNbTaTi lowers N($E_f$) resulting in higher strength of the Mo-heavy alloy. Liu and Shang[34] attribute higher strength of Mo compared to Nb to lower N($E_f$) in Mo.

While DOS is a direct outcome in density functional theory (DFT) calculations that captures the overall bonding environment, extracting bond strength of every bond, especially in HEAs that contain a range of bond chemistries, is not trivial. The knowledge of each bond strength could be very useful to understand the stronger and weaker 'links' in HEAs. In this regard, bond stiffness ($k$) that is readily computed in DFT serves as a good parameter. It is well known that Young's modulus *(E)* can be extracted from the second derivative of energy ($U$) *vs* displacement ($r$), $E \propto \frac{\partial^2 U}{\partial r^2}$. At an equilibrium bond length $r_0$, $E$ is related to bond length as:

$$E = \frac{k}{r_o}, \qquad (1)$$

where $k$ is the bond stiffness. $k$ and $E$ are directly proportional; in our recent works, we showed that $E$ can be simply predicted from $k$ in face-centered cubic (FCC) alloys[35,36]. Higher stiffness translates into stronger bonding, which inherently captures the electronic charge distribution of a bond. Consequently, in this work, we show that $k$ and N($E_f$) are also corelated, thereby leading to a direct correlation between DOS and elastic constants. The significance is that DOS can be used a key descriptor of bond strength, and N($E_f$) can be used to estimate the strength of the materials. Furthermore, because the stiffness of the bonds translates into the resistance to the deformation of the bonds, $k$ is connected to bulk ($B$) and shear ($G$) moduli. We show that $k$ is also a very good descriptor of Pugh ratio ($G/B$), which is a measure of ductility[1,2,37]. This correlation ultimately enables a connection between N($E_f$) and ductility. In addition, in the DFT literature, ductility is also evaluated by the ratio between the surface energy ($\gamma_{surf}$) and unstable stacking fault energy ($\gamma_{USF}$) of specific slip systems, termed as *D parameter*; we show that *D parameter* can also be predicted from N($E_f$) very well, which serves as an independent validation of the correlation between N($E_f$) and ductility. Furthermore, in our recent work[38], we showed that local lattice distortion (LLD) depends on $k$; higher $k$ materials have lower LLD, as stiffer bonds restrict displacement of atoms from their lattice sites. Here, we show that LLD is also fairly corelated to N($E_f$), which then opens a way to connect solid solution strengthening with the electronic structure. In summary, three main RHEA parameters, i.e., elastic bond strength, ductility, and LLD, are shown to be described by DOS.

## 2. Methods

## 2.1. DFT calculations

All DFT are performed using Vienna Ab initio Simulation Package (VASP)[39,40]. All calculations in binary alloys and HEAs are performed in a 3x3x3 body centered cubic (BCC) supercells containing 54 atoms, while ternary alloys calculations are performed in 2x2x3 supercells containing 24 atoms. The chemical randomness in the alloys is modeled using special quasi-random structures (SQS) generated by the Alloy Theoretic Automated Toolkit (ATAT)[41]. For exchange-correlation, the Perdew–Burke–Ernzerhof (PBE)[42] functional, which includes gradient corrections, is employed. The electron-ion interactions are treated using the projector augmented wave (PAW)[43] method. The Brillouin zone is sampled using a 6 × 6 × 6 k-point grid for systems with 54 atoms and an 8 × 8 × 6 grid for those with 24 atoms, both within the Monkhrost-Pack scheme. The energy cutoff and KPOINTS convergence study for 54-atom and 24-atom supercell is shown in Figure S1. A plane wave cutoff of 500 eV is applied for all calculations. The electronic self-consistency energy convergence criterion is $10^{-8}$ eV, while the force convergence criterion for ionic relaxation is set to $-10^{-4}$ eV/Å.

## 2.2. Elastic moduli calculations

The elastic stiffness matrix is determined from relaxed structures using the following stress-strain relationship[44]:

$$\sigma_i = \sum_{i,j=1}^{6} C_{ij}\varepsilon_j \quad (2)$$

where $\sigma_i$ represents the stress corresponding to the strain $\varepsilon_j$, and $C_{ij}$ denotes the elastic tensor of the relaxed structure. The bulk modulus (*B*), Voigt-average shear modulus ($G_V$), and Reuss-average shear modulus ($G_R$) are calculated using the following equations:

$$B = \frac{(C_{11}+C_{22}+C_{33})+2(C_{12}+C_{23}+C_{31})}{9} \quad (3)$$

$$G_V = \frac{(C_{11}+C_{22}+C_{33})-(C_{12}+C_{23}+C_{31})+3(C_{44}+C_{55}+C_{66})}{15} \quad (4)$$

$$G_R = \frac{15}{4(S_{11}+S_{22}+S_{33})-4(S_{12}+S_{23}+S_{31})+3(S_{44}+S_{55}+S_{66})} \quad (5)$$

where $S_{ij}$ are the compliance matrices, the inverse of $C_{ij}$.

The shear modulus ($G$) is the average of $G_V$ and $G_R$:

$$G = \frac{G_V+G_R}{2} \quad (6)$$

The Young's modulus ($E_{VRH}$) is calculated based on the Voigt–Reuss–Hill (VRH) approximation with the following equation:

$$E_{VRH} = \frac{9BG}{3B+G} \tag{7}$$

Table S1 presents optimized lattice parameters, elastic coefficients and Pugh ratio of pure refractory BCC elements, binary, ternary and RHEAs.

### 2.3. Bond stiffness calculations

The supercell approach[45] implemented in ATAT is used to calculate bond stiffnesses. In this approach, the reaction forces on atoms are calculated by perturbing them from their equilibrium positions. These forces and displacements are used to compute the stiffness tensor:

$$F(i) = \mathrm{f}(i,j)u(j) \tag{8}$$

where $F(i)$ is the force acting on $i^{\text{th}}$ atom, $\mathrm{f}(i,j)$ is stiffness matrix, and $u(j)$ is the displacement of $j^{\text{th}}$ atom from its equilibrium. Next, the stiffness matrix is transformed into stretching and bending stiffness model. To ensure the intended transferable properties[46], the transferable force constant technique makes three important assumptions. First, only nearest-neighbor interactions are considered in the stiffness tensor calculations, as these interactions predominantly determine overall bond stiffness. Longer-range force constants are excluded because they have minimal impact on the overall stiffness and are highly dependent on specific local environmental variations. Secondly, to obtain effective isotropic bending stiffnesses, the bending stiffnesses are averaged over different spatial directions. Third, the off-diagonal elements of the bond stiffness tensor are constrained to zero. Consequently, the stretching stiffness (*k*) and the isotropic bending stiffness (*b*) are the only two independent terms remaining in the bond stiffness tensor:

$$\mathrm{f}(i,j) = \begin{pmatrix} b & 0 & 0 \\ 0 & b & 0 \\ 0 & 0 & k \end{pmatrix} \tag{9}$$

Since stretching stiffness is the dominant component that influences bond strength[35,47], and bending component is negligible, bond stiffness is described by stretching term in this work. We calculate the stiffness of the first nearest neighbor (1NN) and second nearest neighbor (2NN) A-A, A-B, and B-B bonds. Finally, the average stiffness ($\overline{k}$) is calculated as the mean of 1NN and 2NN bond stiffness.

### 2.4. LLD calculations

The LLD is determined by calculating the mean atomic displacement of atoms after DFT relaxation, using the standard formula[48] used in literature as:

$$LLD = \frac{1}{N}\sum_i \sqrt{(x_{i-}x_i')^2 + (y_{i-}y_i')^2 + (z_{i-}z_i')^2} \tag{10}$$

where $(x_i, y_i, z_i)$ and $(x_i', y_i', z_i')$ represent the coordinates of atom $i$ in the ideal BCC structure and the relaxed structure, respectively. N denotes to the total number of atoms in the supercell.

## 3. Results

### 3.1. N($E_f$) and mechanical properties

Figure 1 shows the DOS of various alloys including unary, binary, ternary, and quinary HEAs. Figure 1a shows the comparison of DOS between Groups V and VI elements in blue and red colors, respectively. All Group VI elements exhibit lower N($E_f$) values compared to Group V elements and the trend of N($E_f$) is W < Mo < Cr < Ta < Nb < V. While W has the lowest N($E_f$) at 0.45, V has the highest N($E_f$) at 2.15. The difference in N($E_f$) values between Group V and VI is attributed to the *d*-electrons occupancy. Going from Group V to Group VI, the addition of the extra electron fills the lower energy *d*-orbitals. This results in the shift of Femi level towards the pseudo-gap, a valley between bonding and antibonding states, leading to fewer electronic states available at the Fermi level[4,29,32]. Prior studies have also shown that DOS at the Fermi level decreases with the increase in the number of *d* electrons in the transition metal alloys[32,33,49].

The trend persists in alloys. Figure 1(b) shows that as Groups V and IV elements are added to Group VI alloys, the DOS systematically increases. For example, on adding Nb to MoW, N($E_f$) increases from 0.47 to 0.52 in MoWNb. Replacing Group-V Nb with Group-IV Zr, which has one less electron than Nb, raises N($E_f$) to 1.02 in MoWZr. Replacement of Mo with Group IV Ti increases N($E_f$) from 1.02 in WMoZr to 1.41 in WTiZr, illustrating the same trend observed in Figure 1a. The addition of Group IV elements, which have two electrons less than Group VI, further raise N($E_f$) compared to Group V elements.

Figures 1(c) and 1(d) show DOS of RHEAs, i.e., $Mo_xNbTiV_{0.3}Zr$ (where x = 0.1, 0.5, 0.7, 1.0 and 1.3) and $MoNbTaWTi_x$ (where x = 0, 0.25, 0.50, 0.75 and 1), respectively. We specifically choose these two alloys because they have been experimentally investigated previously[50,51]. As Mo concentration is increased in $Mo_xNbTiV_{0.3}Zr$, the N($E_f$) decreases systematically. In contrast, in $MoNbTaWTi_x$, increasing Ti concentration increases N($E_f$), as expected. In the former alloy, addition of a Group VI Mo reduces the concentration of Groups V and IV elements, thereby reducing N($E_f$) whereas opposite happens in the latter alloy with the addition of Group IV Ti. From Figure 1, a general observation is that N($E_f$) decreases from Groups IV towards VI. Table 1 presents the N($E_f$) values for all alloys illustrated in Figure 1. Figure S2 and S3 provide DOS of 25 binary and 15 ternary BCC alloys in this study, and they follow the suggested N($E_f$) trend.

The parameter N($E_f$) is often linked to alloy stability, with lower values of N($E_f$) indicating greater structural stability[26,52]. We analyze the correlation between N($E_f$) and mechanical properties of various alloys. We find that N($E_f$) exhibits a negative correlation with Young's modulus ($E_{VRH}$), as shown in Figure 2(a). The stronger, Group

VI elements are stacked in the top-left corner, whereas the weaker Group V elements are stacked in the bottom right corner of Figure 2(a). Binary alloys follow the same trend. For instance, MoW, consisting of Group VI elements has a higher $E_{VRH}$ value (351 GPa) and a lower N($E_f$) (0.47) compared to NbTi, composed of Group V and IV elements, with a lower $E_{VRH}$ (66 GPa) and a higher N($E_f$) (1.66). The addition of Group IV elements (Ti, Hf or Zr) to Group VI alloy lowers $E_{VRH}$ and raises N($E_f$) much more than the addition Group V (V, Ta or Nb) elements. For example, ternary alloy MoWTa has a higher $E_{VRH}$ (287 GPa) than MoWTi $E_{VRH}$ (203 GPa), whereas its N($E_f$) is 0.54, which lower than that in MoWTi, which is 1.1. Similarly, in the two HEAs, i.e., $Mo_XNbTiV_{0.3}Zr$ and $MoNbTaWTi_X$, $E_{VRH}$ increases while N($E_f$) decreases as Mo content increases in the former, whereas $E_{VRH}$ decreases and N($E_f$) increases with higher Ti content in the latter, indicating that both alloys follow the elastic strength *vs* N($E_f$) correlation.

Since $E_{VRH}$ and $\bar{k}$ are corelated, a correlation between N($E_f$) and $\bar{k}$ is now unsurprising; this correlation is shown in Figure 2(b). Because Group VI elements are elastically stronger, their bond stiffness are higher as well, compared to Group V and IV elements. For example, a ternary alloy MoWNb has higher $\bar{k}$ (2.55) but lower N($E_f$) (0.52) compared to MoWZr, which has lower $\bar{k}$ (1.89) but higher N($E_f$) (1.02). In $Mo_XNbTiV_{0.3}Zr$ HEA, the $\bar{k}$ increases but N($E_f$) decreases with the increase in Mo content, as expected. Conversely, in $MoNbTaWTi_X$, the $\bar{k}$ decreases but N($E_f$) increases with the increase in Ti content. These trends indicate that the bond stiffness captures the effect of bonding strength per the filling of the *d*-electrons, i.e., as the number of *d*-electrons increase from Group IV to VI, the overall bonding strength increases. As expected, Figure 2(c) demonstrates a strong positive correlation between $\bar{k}$ and $E_{VRH}$. The alloy with stiffer bonds possesses a higher Young's modulus and vice versa.

The macroscopic properties such as bulk and shear moduli are the consequence of the atomic level bond strengths and electronic interactions. Recent studies have shown that bond stiffness can be correlated with bulk modulus[53,54]. Therefore, it is reasonable to investigate the correlation between $\bar{k}$, *B,* and *G* in RHEAs. Figure 3(a) reveals a correlation between $\bar{k}$ and *B*. This correlation is expected because materials with stiffer bonds exhibit higher resistance to compression. Group VI elements (Cr, Mo, and W) exhibit higher values of $\bar{k}$, and *B* compared to Group V elements (V, Nb, and Ta). Additionally, the binary and ternary alloys containing Group VI elements have higher $\bar{k}$ and *B.* For example, CrW has a higher $\bar{k}$ (2.63 eV Å$^{-2}$) and *B* (279 GPa) than TaTi, which has a lower $\bar{k}$ (0.82 eV Å$^{-2}$) and *B* (144 GPa). Similarly, the ternary alloy MoWTa has a higher $\bar{k}$ (2.67 eV Å$^{-2}$) and *B* (253 GPa) compared to WTiZr, which has lower $\bar{k}$ (1.15 eV Å$^{-2}$) and *B* (127 GPa). In case of $Mo_XNbTiV_{0.3}Zr$, $\bar{k}$ *and B* both increase with the increase in Mo content whereas in $MoNbTaWTi_X$, and $\bar{k}$ and *B* both decrease with the increase in Ti content, as shown in Figure 3(a). Since the material deformation under shear stress is also closely related to the bond stiffness, we also find correlation between $\bar{k}$ and *G,* as shown in Figure 3(b). By extension, the Pugh ratio, (*G*/*B*), which is an indicator of ductility also corelates with $k$, as shown in Figure 3(c). A lower Pugh ratio typically indicates tendency of a ductile failure. The alloys that have lower values of $\bar{k}$ are more ductile, which is expected because the less stiff bonds are more stretchable. A comprehensive database of $E_{VRH}$, $B$, $G$, $\overline{k}$ and Pugh ratio for pure elements, binary, ternary and quinary HEAs are presented in Table S1.

To gain insight into the role of electronic structure on the ductility of a material, we explore the correlation between N($E_f$) and Pugh ratio, as shown in Figure 4. A very good correlation is observed, where the alloys with lower N($E_f$) values demonstrate higher Pugh ratios. Group VI elements, as well as the alloys that consist of a higher concentration of Group VI elements, exhibit lower N($E_f$) and higher Pugh ratios. In contrast, Group V elements and their alloys have higher N($E_f$) and lower Pugh ratios. For example, the binary alloy MoW possesses the lowest N($E_f$) and highest Pugh ratio compared to the other binary alloys. Furthermore, the addition of Group IV elements such as Ti or Zr leads to a significant decrease in the Pugh ratio and an increase in N($E_f$) compared to the addition of Group V elements such as Nb or Ta. This correlation generally holds good for HEAs as well. The Pugh ratio increases while N($E_f$) decreases with the increase in Mo content in $Mo_XNbTiV_{0.3}Zr$, whereas the decrease in Pugh ratio and the increase in N($E_f$) is observed in $MoNbTaWTi_X$ with increase in Ti content.

To understand the effect of electronic charge distribution on the material strength, we analyze the charge density of MoW and MoWTi along the (110) plane as shown in Figures 5a and 5b. The charge density plots of MoW and MoWTi reveal significant differences in bonding characteristics. In MoW, the charge density is more localized and directional between Mo and W atoms, indicating stronger *d*–*d* hybridization and covalent character. This results in stiffer bonds and enhanced bond strength. In contrast, the addition of Ti in MoWTi leads to a more delocalized and isotropic charge density, particularly around Ti atoms, suggesting weaker directional bonding. The reduced charge accumulation between Mo, W, and Ti atoms implies a weakening of *d*–*d* interactions, thereby increasing the metallic character of the alloy. This aligns with the observed increase in N($E_f$) and the shift toward a more free-electron-like metallic behavior, ultimately leading to reduced bond stiffness.

### 3.2. N($E_f$) and LLD

In our recent work[38], we observed that alloys with higher bond stiffness tend to have lower lattice distortion (LLD). This is because the atoms that are bonded via stiffer bonds are tightly held together making them less prone to displacements from their lattice sites. Here, we investigate the correlation between N($E_f$) and LLD. Figure 6(a) shows that although the correlation between LLD and N($E_f$) is not particularly strong, there is still a general trend where LLD increases with N($E_f$). We find that alloys composed of Group IV elements (e.g., Ti, Hf, Zr) have relatively higher LLD compared to other alloys especially those containing Group VI elements. This suggests that additional factors likely contribute to LLD beyond just electronic effects. A key factor influencing this behavior is the atomic radius mismatch. Hf (1.59 Å) and Zr (1.60 Å) have significantly larger metallic radii compared to Group VI elements such as Cr (1.29 Å), Mo (1.39 Å), and W (1.39 Å), as well as Group V elements like V (1.34 Å), Nb (1.46 Å), and Ta (1.46 Å). This larger atomic size mismatch leads to greater LLD, as the surrounding atoms experience significant geometric constraints in accommodating the size differences. Thus, while $N(E_f)$ influences LLD via bond stiffness, the pronounced distortion in Hf- and Zr-containing alloys is also driven by their larger atomic radii. The effect of atomic radii on LLD has been previously documented in literature[15,55–57]. We further observe that the correlation between $N(E_f)$ and LLD becomes stronger within a given alloy, such as $Mo_XNbTiV_{0.3}Zr$ and $MoNbTaWTi_X$, as shown in Figures 6(b) and 6(c), respectively. Tandoc et al.[1] recently developed a surrogate model essentially based on $N(E_f)$ illustrating that LLD can be effectively predicted from the filling of the *d*-band. They also found that LLD has a good correlation with VEC in simpler binary and ternary alloys. From Figure 2, because $N(E_f)$ and $k$ are connected, Figure 6 further indicates that filling of the *d*-band directly influences both $N(E_f)$ and LLD. However, the weaker correlation observed in Figure 6(a) highlights that while electronic effects $N(E_f)$ contribute to LLD, additional structural factors, particularly atomic radius mismatch play a crucial role in increasing lattice distortions in alloys containing Hf and Zr.

## 4. Discussion

In literature, the DOS at the Fermi level has been linked to alloy stability. Xu et al.[26] show that energetically favorable atomic arrangements in intermetallics often correspond to low $N(E_f)$ values. Soderlind et al.[27] showed that the stability of crystal structures is influenced by the *d*-electron DOS, where differences in *d*-band filling affect whether a metal stabilizes in a BCC structure or close-packed structures like FCC or hexagonal closed packed (HCP). Similarly, Iotova et al.[28] demonstrate that a lower DOS at the Fermi level, as observed in $Ni_3Si$, correlates with stronger bonding and greater stability, while a higher DOS in $Ni_3Mn$, suggests weaker bonding and lower stability due to increased metallic character. San et al.[49] reveal that alloys with a higher proportion of Group VI elements, which exhibit lower DOS at the Fermi level, are more stable compared to alloys rich in Group IV or V elements. In line with these previous findings, our results further connect $N(E_f)$ with various mechanical properties, establishing it as a key predictor of bond strength, stiffness, and ductility in refractory alloys.

VEC, which accounts for the total number of valence electrons in the alloy composition, including *d*-electrons in the valence band, is widely used to predict the mechanical properties and phase stability of HEAs. Ikehata et al.[58] demonstrated a linear correlation between VEC and elastic constants ($C_{11}$ and $C_{12}$) in binary Ti and Zr alloys. Sorkin et al.[59] observed that $E$ tends to increase with higher VEC. However, there is a caveat attached to VEC. Since VEC is simply a numerical value that provides a general measure of number of electrons, multiple alloys can share the same VEC while exhibiting different $E$ values. The results by Sorkin et al.[59] show a wide variation in E for a given VEC in multiple HEAs. Our findings extend the current understanding by suggesting that, while VEC has been used in predicting phase stability, it has limitations when assessing mechanical properties due to less specific electronic interpretation as illustrated in Figure 7. Figures 7(a) and 7(c) show that multiple alloys with varying $E_{VRH}$ and Pugh ratios share the same VEC. In contrast, in Figure 7(b) and 7(d), a better correlation is observed between these mechanical properties and $N(E_f)$. Note that DOS incorporates the real electronic environment after mixing of the elements into their relaxed nearest neighbor environments.

The *D parameter* has been suggested as another measure of ductility[1,2,37]. An alloy with higher *D parameter* is considered to have a greater likelihood of exhibiting intrinsic ductility. Here, we connect the *D parameter* with N($E_f$). We take the data of *D parameter* of a large number of alloys from the work of Hu et al.[37] However, since their data does not include the corresponding N($E_f$) values, we predict N($E_f$) using the following steps. We use Hu et al.[37] alloy compositions to calculate $E_{VRH}$ using rule of mixtures (ROM). Figure S4 shows a good correlation between the DFT-calculated $E_{VRH}$ and ROM $E_{VRH}$ values of some alloys. Next, we develop a linear regression model between $E_{VRH}$ and N($E_f$) of various alloys that are shown in Figure 2(a). Subsequently, the model is used to predict N($E_f$) for the Hu et al.[37] alloys. Notably, a good correlation between predicted N($E_f$) and the *D parameter* is obtained, as shown in Figure 8(a). These findings suggest that not only N($E_f$) can be predicted from $E_{VRH,}$ but the *D parameter* can also be estimated from N($E_f$). This correlation also provides an independent validation of the correlation between N($E_f$) and ductility presented above in the form of Pugh ratio. We also examined the correlation between two ductility parameters, the Pugh ratio and the *D parameter*. Using a linear regression model based on N($E_f$), we predicted the *D parameter* for the alloys shown in Figure 4. Figure 8(b) demonstrates a good correlation between these two parameters, illustrating that both lead to converged qualitative trends regarding the ductility of the alloys.

While this work focuses on the properties of disordered alloys, it is worth commenting on the potential impact of short-range order (SRO) on the electronic structure and mechanical properties of alloys which have been under intense spotlight[4,60,61]. To explore this, we compare the electronic and mechanical properties of some ordered (B2) and disordered (SQS) binary alloys. As shown in the Table 2, the variation in $E_{VRH}$ and N($E_f$) between the B2 and SQS structures differs among alloy systems. For CrW and MoW, N($E_f$) remains nearly unchanged, suggesting minimal impact of ordering on the electronic states at the Fermi level. In contrast, WTa and VNb exhibit significant reductions in N($E_f$) upon disordering. This difference is likely due to the different number of *d*-electrons in W and Ta (and V and Nb), which affect bonding depending on the order or disorder. In contrast, the number of *d*-electrons are same in Cr and W and (Mo and W). This analysis indicates that structural order or disorder (and SRO) could influences bond strength and mechanical properties.

## 5. Conclusion

This paper investigates the role of electronic density of states at the Fermi level, N($E_f$), as a predictor of mechanical properties in RHEAs. Our findings reveal linear correlations between N($E_f$) and key mechanical properties, i.e., elastic moduli, bond stiffness, Pugh ratio, and LLD. Lower values of N($E_f$) correspond to stronger bonds, higher Young's modulus, increased stiffness, and decreased LLD. Our calculations show that Group VI elements that have additional *d* electrons lead to a lower N($E_f$) due to more filled *d*-band. In contrast, Group IV elements, that consist of two less electrons, have higher

N($E_f$). This difference significantly influences alloy characteristics, including bond stiffness and ductility. The alloys with higher content of Group VI elements have relatively lower N($E_f$) compared to the alloys that have higher concentrations of group V or Group IV elements.

While VEC has been widely used to predict phase stability and mechanical trends in high-entropy alloys (HEAs), our study highlights its limitations in evaluating mechanical properties. Specifically, VEC provides only a numerical count of valence electrons, without fully capturing the complex electronic structure variations among different alloys. As a result, multiple alloys may share the same VEC yet exhibit significant differences in elastic modulus, stiffness, and ductility. Our results emphasize that N($E_f$), which directly incorporates detailed electronic structure information from DOS calculations, serves as a more precise and reliable descriptor for predicting mechanical behavior.

Additionally, we show that N($E_f$) can be predicted from ROM $E_{VRH}$ via a linear regression model, which ultimately enables a correlation between N($E_f$) and intrinsic ductility, i.e., the *D parameter*. This correlation demonstrates that both N($E_f$) and *D parameter* can be estimated from the elastic modulus, offering valuable insights for alloy design. In summary, our findings show that N($E_f$) is a reliable descriptor of bond stiffness (and hence elastic constants), ductility (i.e., Pugh ratio and *D parameter*), and LLD.

## Author Contributions

**D. Pant**: DFT calculations (lead); Formal analysis (equal); Methodology (equal); Visualization (lead); Writing (equal). **D. S. Aidhy**: Conceptualization; Formal analysis (equal); Methodology (equal); Project administration (lead); Supervision (lead); Writing – review & editing.

## Acknowledgments

This work is supported by the US Department of Energy, Office of Science, Basic Energy Science, Mechanical Behavior and Radiation Effects program.

## Data availability

All data supporting the findings of this study are available within the article and its Supplementary Information. All other data are available from the corresponding author upon request.

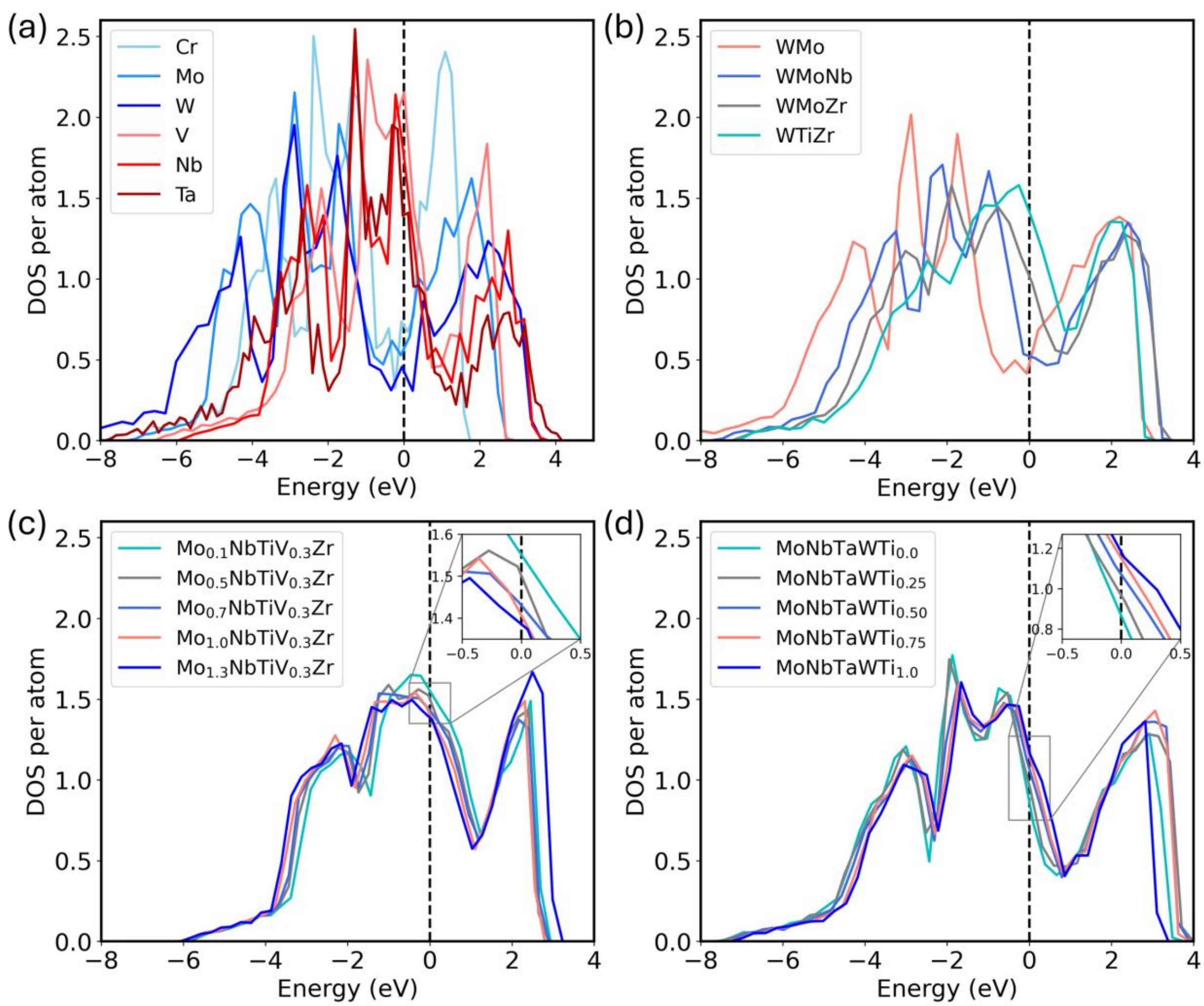

**Figure 1. DOS per atom of unary, binary, ternary, and HEAs.** (a) DOS of Group V elements (V, Nb and Ta) and Group VI elements (Cr, Mo, and W). (b) DOS of W-alloys. (c) DOS of $Mo_XNbTiV_{0.3}Zr$ alloys, where x= 0.1, 0.5, 0.7, 1, 1.3. (d) DOS of $MoNbTaWTi_X$ alloys, where x= 0, 0.25, 0.50, 0.75, 1. The dotted line at 0 eV indicates the Fermi level.

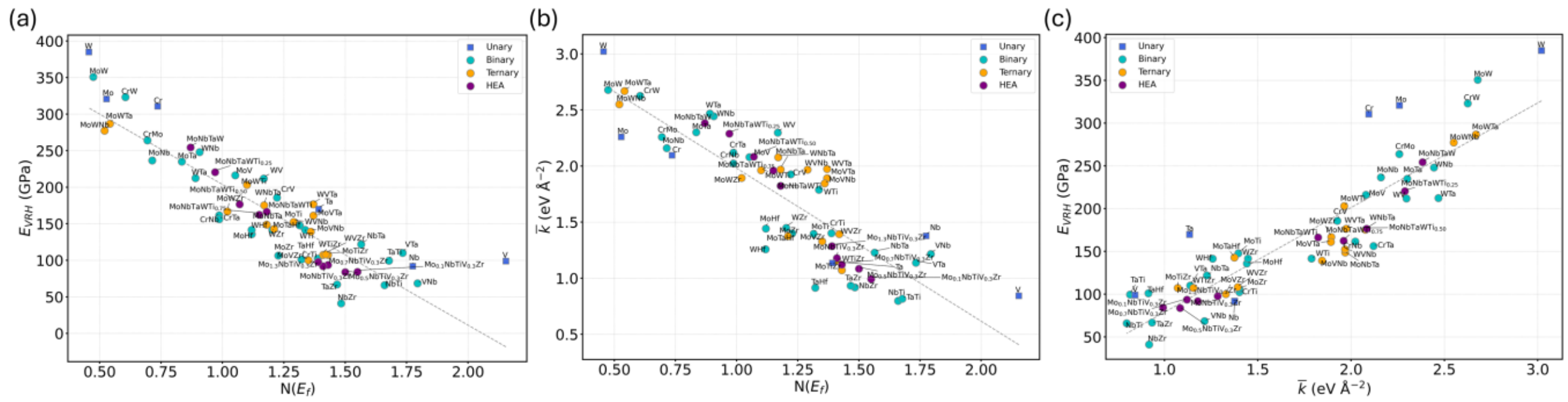

**Figure 2.** Correlation between (a) N($E_f$) and $E_{VRH}$, (b) N($E_f$) and $\overline{k}$, and (c) $\overline{k}$ and $E_{VRH}$ in various alloys.

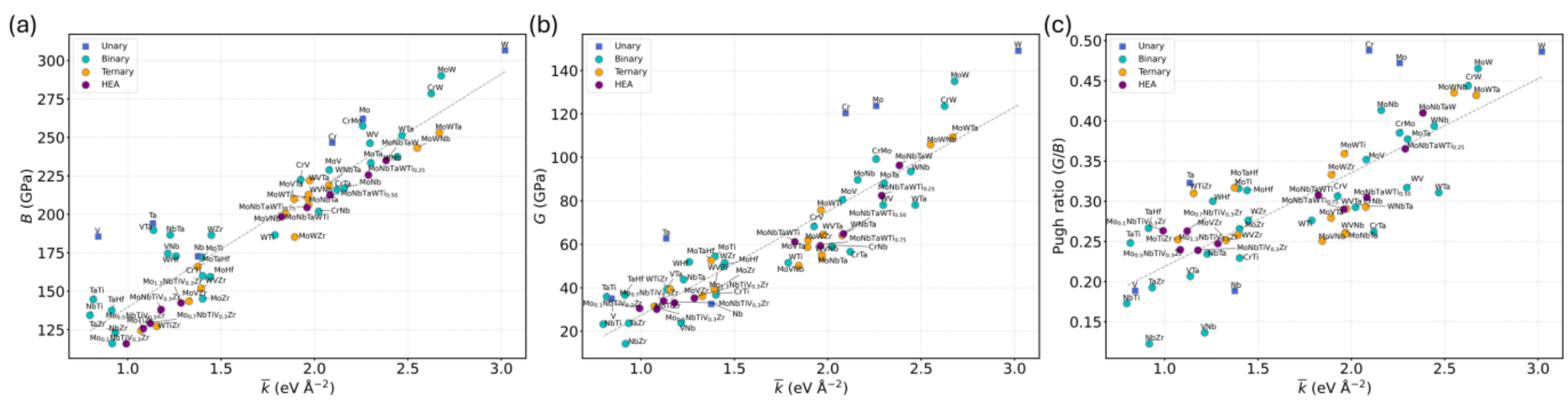

**Figure 3.** Correlation between (a) $\overline{k}$ and $B$, (b) $\overline{k}$ and $G$, and (c) $\overline{k}$ and Pugh ratio in various alloys.

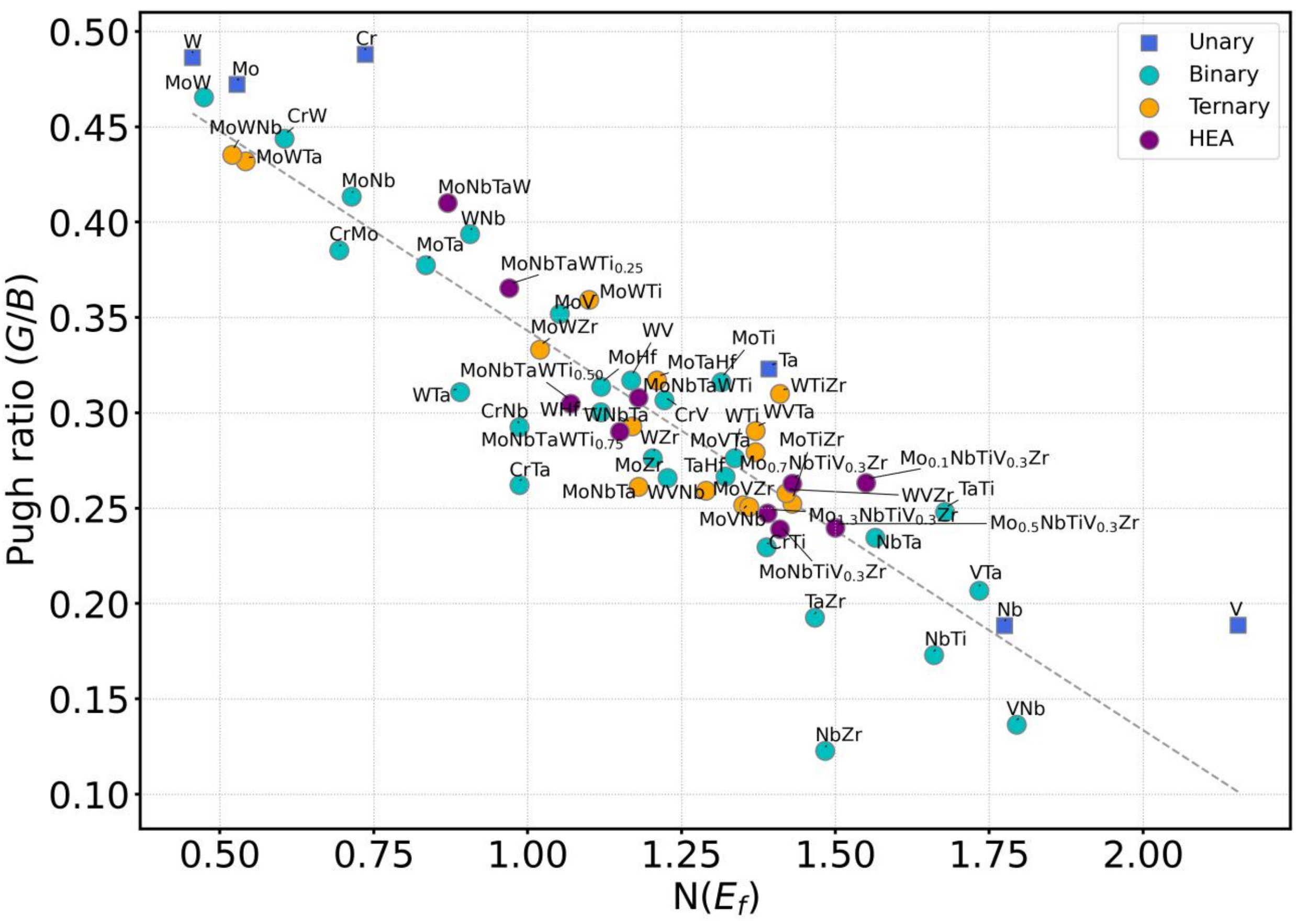

**Figure 4.** Correlation between N($E_f$) and Pugh ratio ($G/B$) in various alloys.

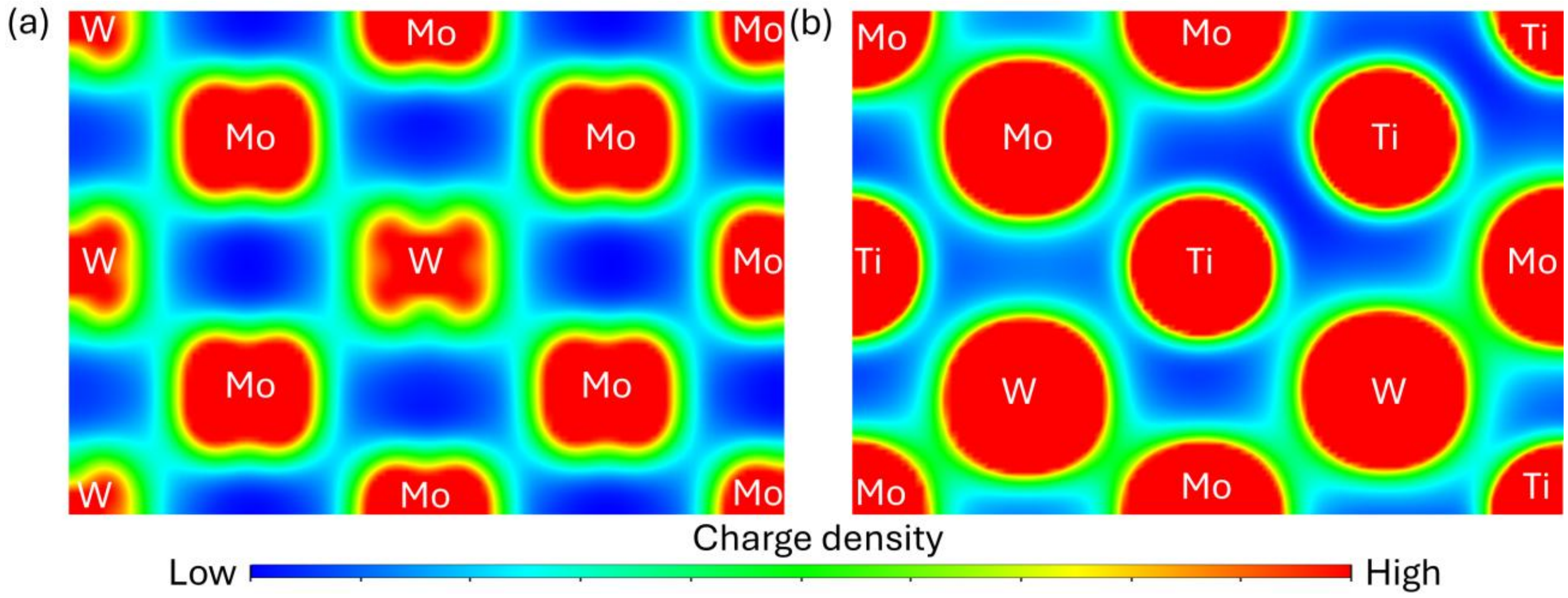

**Figure 5.** Charge density distribution along the (110) plane for (a) MoW and (b) MoWTi.

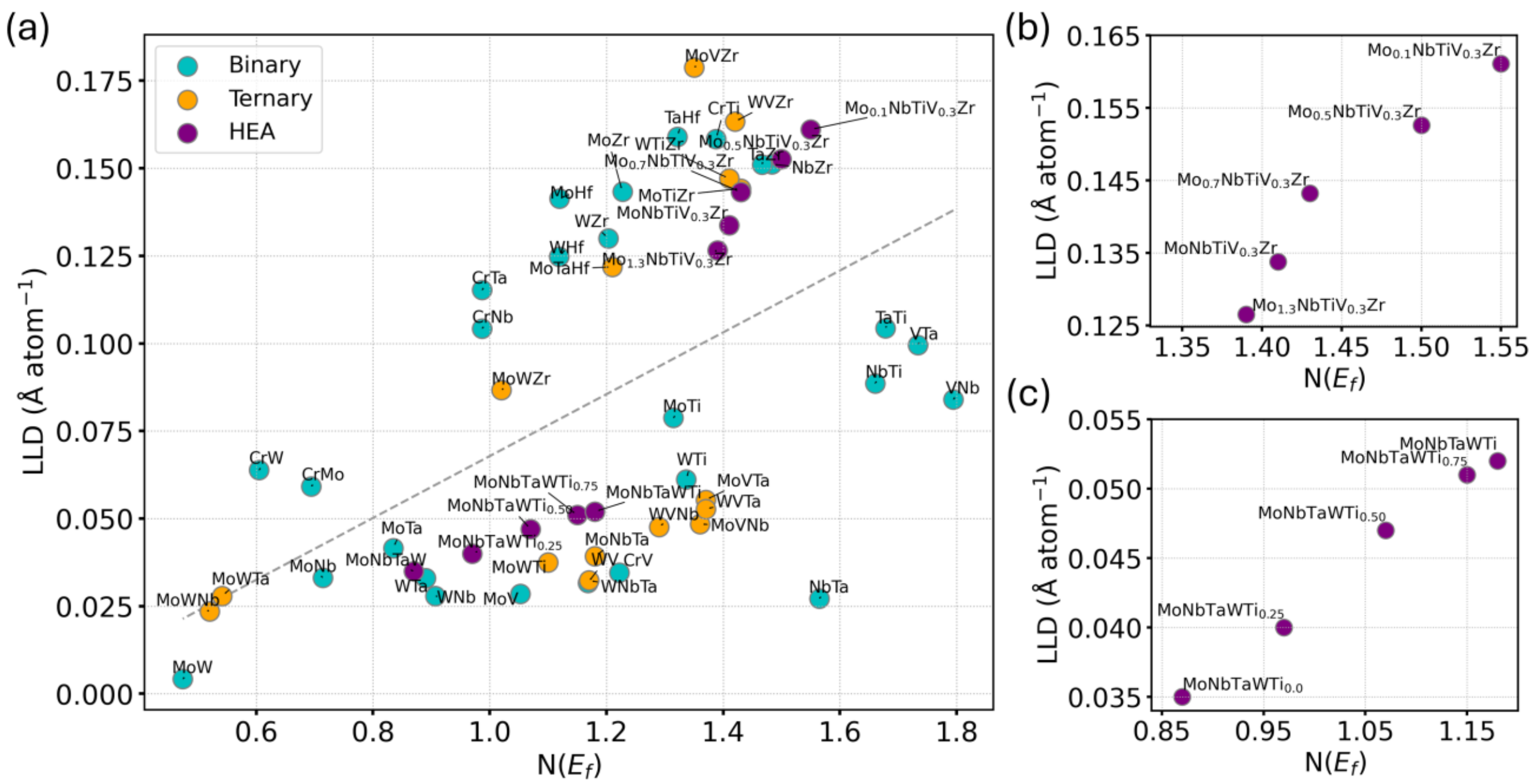

**Figure 6.** Correlation between N($E_f$) and LLD in (a) various alloys, (b) in $MoXNbTiV_{0.3}Zr$ alloys, where x = 0.1, 0.5, 0.7, 1, 1.3. (c) in $MoNbTaWTi_X$ alloys, where x = 0, 0.25, 0.50, 0.75. LLD increases with the increase in N($E_f$).

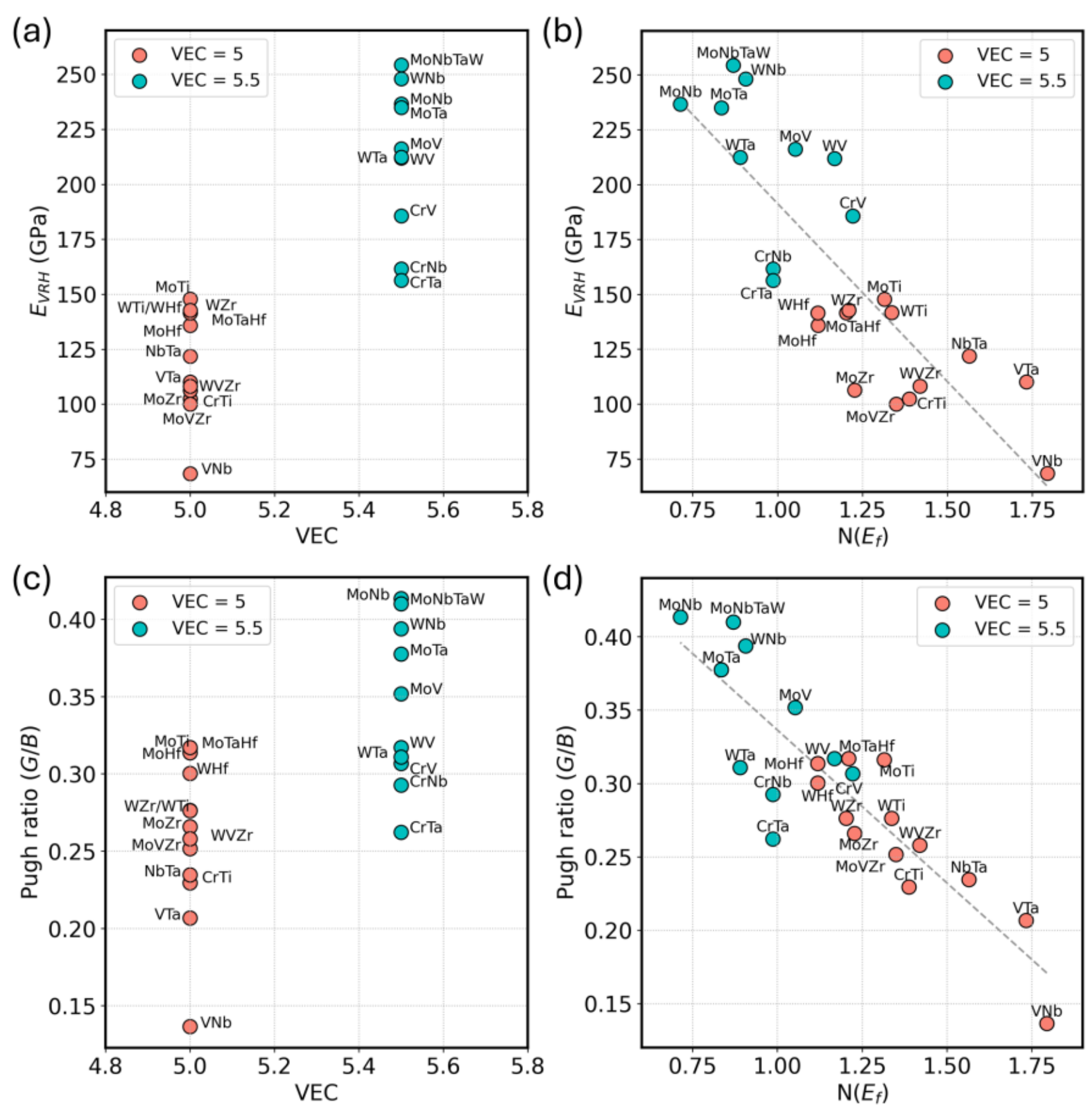

**Figure 7. Comparison between VEC *vs* N($E_f$) in predicting $E_{VRH}$ and Pugh ratio.** (a), (c) Correlation of VEC with $E_{VRH}$ and Pugh ratio. (b), (d) Correlation of N($E_f$) with $E_{VRH}$ and Pugh ratio.

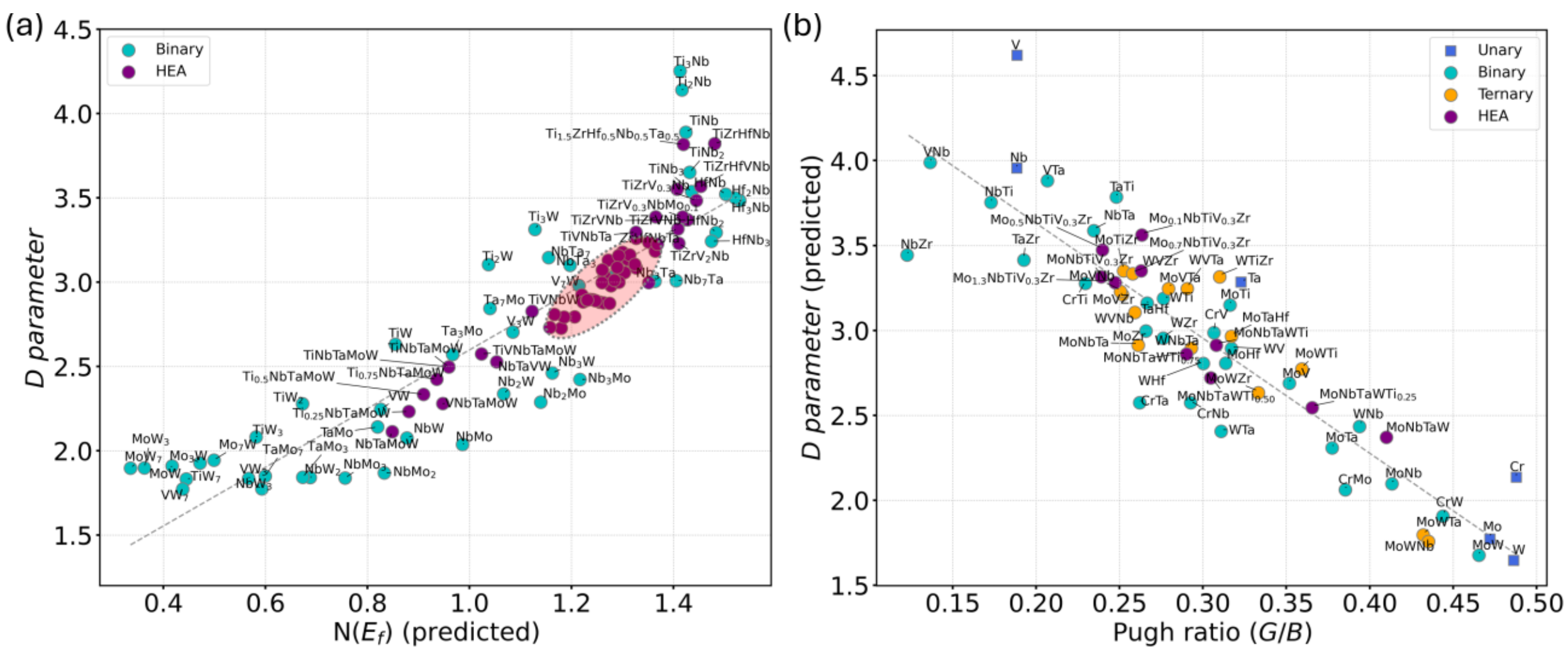

**Figure 8.** (a) Correlation between predicted N($E_f$) and *D parameter* [37] in various alloys. HEAs highlighted in red ellipse contain Ti, V, Zr Nb and Mo elements. For clarity, the highlighted alloys are separately plotted in Figure S5 (Supplementary Information). (b) Correlation between Pugh ratio and predicted *D parameter*.

**Table 1.** N($E_f$) of unary, binary, ternary, quaternary and quinary HEAs.

| **Composition** | **N($E_f$)** | **Composition** | **N($E_f$)** |
|---|---|---|---|
| Cr | 0.74 | MoVZr | 1.35 |
| Mo | 0.53 | MoTiZr | 1.43 |
| W | 0.45 | $Mo_{0.1}NbTiV_{0.3}Zr$ | 1.55 |
| V | 2.15 | $Mo_{0.5}NbTiV_{0.3}Zr$ | 1.50 |
| Nb | 1.78 | $Mo_{0.7}NbTiV_{0.3}Zr$ | 1.43 |
| Ta | 1.39 | $Mo_{1.0}NbTiV_{0.3}Zr$ | 1.40 |
| WMo | 0.47 | $Mo_{1.3}NbTiV_{0.3}Zr$ | 1.38 |
| WMoNb | 0.52 | $MoNbTaWTi_{0.0}$ | 0.87 |
| WMoZr | 1.02 | $MoNbTaWTi_{0.25}$ | 0.97 |
| WTiZr | 1.41 | $MoNbTaWTi_{0.50}$ | 1.07 |
| MoTa | 0.83 | $MoNbTaWTi_{0.75}$ | 1.15 |
| MoNbTa | 1.18 | $MoNbTaWTi_{1.0}$ | 1.18 |

**Table 2.** Comparison of electronic density of states at the Fermi level N$(E_f)$ and Young's modulus ($E_{VRH}$) for ordered (B2) and disordered (SQS) structures across different binary alloy systems.

| **Alloy** | **N($E_f$)** | | **$E_{VRH}$ (GPa)** | |
|---|---|---|---|---|
| | **B2** | **SQS** | **B2** | **SQS** |
| CrW | 0.59 | 0.60 | 325.83 | 323.12 |
| MoW | 0.49 | 0.47 | 325.56 | 350.72 |
| WTa | 0.98 | 0.89 | 160.49 | 212.35 |
| VNb | 2.28 | 1.79 | 44.35 | 68.41 |